\pdfoutput=1 
\documentclass[10pt, conference,onecolumn]{IEEEtran}

\usepackage{etoolbox}
\patchcmd{\section}{\centering}{}{}{}

\usepackage{titlesec}
\titlespacing*{\section}{0pt}{*2}{*1}
\titlespacing*{\subsection}{0pt}{*3}{*1}

\usepackage{amsmath, amsfonts}
\usepackage[text={5.5in,9.3in}]{geometry}

\usepackage{graphicx}
\usepackage[labelfont=bf]{caption}
\usepackage{subcaption}
\usepackage{colortbl}
\usepackage{nicematrix}

\usepackage[hidelinks,bookmarks=true]{hyperref}
\usepackage[numbered]{bookmark}

\usepackage{listings}
\usepackage[numbers,square,sort&compress]{natbib}

\hypersetup{pdfinfo={
  Author={Eric Leu},
  Title={Fast Consistent Hashing in Constant Time},
}}

\title{\textbf{\\ \LARGE Fast Consistent Hashing in Constant Time}
\vspace{0.2em}}
\author{\large Eric Leu\\
\smallskip
\emph{\normalsize PayPal}}
	
\begin{document}
\maketitle
\thispagestyle{plain}
\pagestyle{plain}
\noindent
\section*{\pdfbookmark[1]{Abstract}{A}\textbf{Abstract}}
\medskip
\noindent
Consistent hashing is a technique that can minimize key remapping when the number of hash buckets changes. The paper proposes a fast consistent hash algorithm (called \emph{power consistent hash}) that has $O(1)$ expected time for key lookup, independent of the number of buckets.  Hash values are computed in real time.  No search data structure is constructed to store bucket ranges or key mappings.   The algorithm has a lightweight design using $O(1)$ space with superior scalability. In particular, it uses two auxiliary hash functions to achieve distribution uniformity and $O(1)$ expected time for key lookup. Furthermore, it performs consistent hashing such that only a minimal number of keys are remapped when the number of buckets changes. Consistent hashing has a wide range of use cases, including load balancing, distributed caching, and distributed key-value stores.  The proposed algorithm is faster than well-known consistent hash algorithms with $O(\log n)$ lookup time. \\
\section*{\pdfbookmark[1]{Introduction}{S1}\textbf{I.\quad Introduction}}
\medskip
\noindent
The concept of \emph{consistent hashing} was introduced by Karger et al. \cite{KL}.  There are two core requirements for consistent hashing: (i) keys are uniformly distributed among buckets; and (ii) key remapping is minimized when the hash space changes. These are referred to as \emph{balance} and \emph{monotonicity} respectively in \cite{KL}.  It is desirable to have an algorithm that can meet the two requirements and also compute hash values mathematically, rather than store bucket ranges or boundaries in a 
data structure that requires $O(n)$ space and $O(\log n)$ time for lookup. With that objective, the proposed 
consistent hash algorithm is unique such that lookup takes only $O(1)$ space and $O(1)$ expected time, which is the same as in ordinary hashing.\\ \\
Consistent hashing is more advanced than ordinary hashing, which uses modular arithmetic to compute the hash value for a given hash key such as $x = key$ mod $n$, 
where $n$ is the number of buckets available for assignment and $x$ is the assigned bucket number.  The computation takes $O(1)$ time.  However, changing the number of buckets will force remapping most of keys to different buckets.  The reshuffling is undesirable due to data redistribution cost and system disruption. On the other hand, consistent hashing can minimize key remapping but typically at the cost of increased lookup time greater than $O(1)$. \\ \\
Compared to ordinary hashing technique, consistent hashing is a more suitable technique for the scenario that involves a changing number of buckets.  It can minimize the number of keys that need to be remapped.  For example, suppose that there are $n$ buckets and $K$ keys.  Each key is mapped to one of the buckets.  Suppose further that after two new buckets are added, a total of 10 million keys are remapped to the two new buckets. Then, under the constraint of consistent hashing, the total number of remapped keys is exactly 10 million.  Each of  the remaining $K-10$ million keys must be mapped to the same bucket before and after the addition of the two buckets.\\ \\
Without loss of generality, the term bucket in the discussion may represent a resource or storage node in a cluster. Multiple keys can be mapped to a single bucket.  Within the bucket, records are searchable by the underlying storage system.  In this paper, key lookup refers to the operation to find the corresponding bucket for a given key, excluding the search of the record within the bucket. \\ \\
\noindent
\textbf{Major Contributions.} \enspace The power consistent hash (Power CH) algorithm takes a unique approach by using two auxiliary hash functions \cite{LE} to accomplish three important objectives: 
\begin{itemize}
\smallskip
\item Distribute keys among buckets with equal probability.  This property holds even after the number of buckets changes. 
\smallskip
\item Minimize key remapping when the number of buckets changes. This reduces data redistribution cost and system disruption. 
\smallskip
\item Perform key lookup in $O(1)$ space and $O(1)$ expected time. 
\end{itemize}
\ \\
Unlike some other algorithms \cite{AO,EY,KL,TR,WR}, the algorithm does not construct  any dynamic data structure that requires update due to bucket removal or addition. No extra computation cost is involved.  System disruption is further reduced.\\  \\
\textbf{Related Work.} \enspace There are different algorithms to implement consistent hashing.  In Karger et al. \cite{KL}, a ring hash algorithm divides a unit circle into bucket ranges.  It constructs a data structure to store bucket ranges and supports key lookup by search.  To distribute keys more evenly among buckets, it needs to distribute a considerable number of \emph{points} on the unit circle to further divide each bucket into multiple virtual nodes.  The design increases lookup time and memory footprint, causing scalability issues in dealing with a large number of buckets.  The algorithm also needs to update the data structure to support bucket removal or addition, which will impact system availability.  Other well-known consistent hash algorithms include rendezvous hashing \cite{TR,WR} and jump consistent hash \cite{LV}.  The lowest lookup time among those \cite{KL,LV,TR,WR} and variants (such as \cite{AO}) is $O(\log n)$, where $n$ is the number of buckets.  The hash algorithms \cite{AO,KL,TR,WR} also use $O(n)$ space, resulting in a lower scalability.  Additionally, the hash algorithm in Maglev \cite{EY} is a specialized solution for a particular use case and does not ensure \emph{monotonicity} defined in \cite{KL}. It also constructs a lookup table of size $kn$ (typically $100n$ or more), for which it takes a long time to initialize or update. Scalability is significantly reduced.\\ \\ 
The rest of the paper is organized as follows.  Section II  demonstrates the algorithm with an example.  Section III constructs the main hash function based on two auxiliary hash functions.  Section IV describes probability distributions of hash values.  One auxiliary function produces uniformly distributed hash values.  The other auxiliary function returns hash values that have a weighted probability distribution.  Section V defines mapping consistency for which key remapping is minimized when the number of buckets changes.  Section VI presents algorithms to compute the two auxiliary functions.  Section VII provides performance test results.  Section VIII describes a rehashing technique with unique advantages. Section IX concludes the paper. 
\begin{figure}[t!]
\centering
\begin{subfigure}{1\textwidth}
\centering
\includegraphics[scale=1.0]{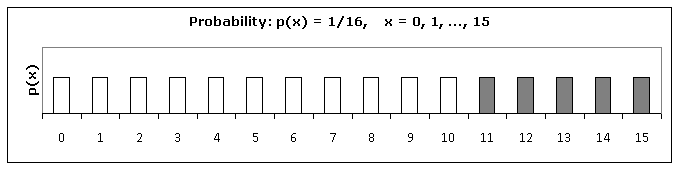}
\caption{Step 1: Apply $f(key, 16)$}
\end{subfigure}
\par\bigskip
\begin{subfigure}{1\textwidth}
\centering
\includegraphics[scale=1.0]{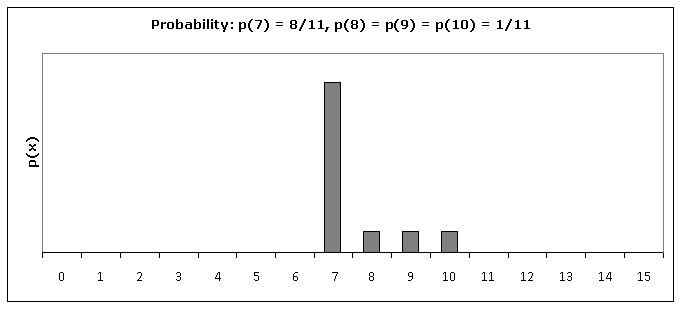}
\caption{Step 2: Apply $g(key, 11, 7)$}
\end{subfigure}
\par\bigskip
\begin{subfigure}{1\textwidth}
\centering
\includegraphics[scale=1.0]{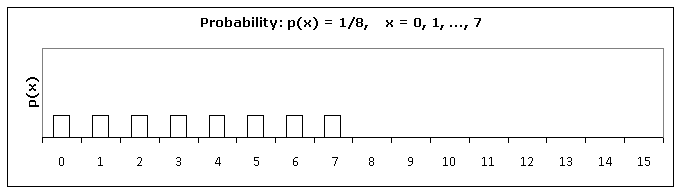}
\caption{Step 3: Apply $f(key, 8)$}
\end{subfigure}
\caption*{\textbf{Figure 1 } \enspace Example to Compute $hash(key, 11)$}
\end{figure}
\medskip
\section*{\pdfbookmark[1]{An Example}{S2}\textbf{II.\quad An Example}}
\medskip
\noindent
Figure 1 illustrates how the power consistent hash algorithm computes $hash(key, 11)$ to produce an integer $x \in \{0, 1, \ldots, 10\}$ with equal probability for all possible $key$. It is done in such a way that the other two objectives are also achieved, namely mapping consistency (defined later in Section V) and $O(1)$ expected running time.  \\ \\
In the example, $hash(key, 11)$ is computed using two auxiliary hash functions $f()$ and $g()$, which takes up to three steps as follows. Figure 1 shows the probability distributions of hash values in each step.  In Step 1, $f(key, 16)$ is applied to get the first result $x$, where the number 16 is chosen because 16 is the smallest power of 2 greater than 11.  
\medskip 
\begin{description}
\item[$\textbf{Step 1:}$] \quad Apply $f(key, 16)$ and check the integer $x$ returned.  The function call $f(key, 16)$ maps $key$ to an integer $x \in \{0,1,\ldots,15\}$ with equal probability.  If $x$ is in the range $[0,10]$, the value of $x$ is the result.  Otherwise, $x$ is in the range $[11, 15]$.  Go to Step 2 to remap $key$ to an integer in the range $[0, 10]$. 
\medskip
\item[$\textbf{Step 2:}$] \quad Apply $g(key, 11, 7)$ and check the integer $x$ returned.  The function call $g(key, 11, 7)$ maps $key$ to an integer $x \in \{7,8, 9,10\}$ with a weighted probability as shown in Figure 1(b), where probability $p(7)=8/11$, and $p(8)=p(9)=p(10)=1/11$.  If $x \in \{8,9,10\}$, the value of $x$ is the result.  Otherwise, for $x=7$,  proceed to Step 3.
\medskip
\item[$\textbf{Step 3:}$] \quad Apply $f(key, 8)$ to get the result $x$.  The function call $f(key, 8)$ maps $key$ to an integer $x \in \{0, 1,\ldots,7\}$ with equal probability as shown in Figure 1(c).  The value of $x$ is the result.
\\
\end{description} 
In Step 1, $f(key, 16)$ maps $key$ to an integer $x$ in the range $[0, 15]$ with equal probability.  If $x$ returned from $f(key, 16)$ is outside the valid range $[0, 10]$, Steps 2 and 3 together are used to remap $key$ to an integer in the range $[0, 10]$ with equal probability as follows. First, in Step 2, $g(key, 11, 7)$ maps $key$ to an integer $x \in \{7,8,9,10\}$ with a weighted probability given by 
\[
P(X=x) =
\begin{cases}
\dfrac{8}{11}, &
x = 7, \\ \\
\dfrac{1}{11}, &
x = 8, 9, 10.
\end{cases}
\]
\\
If $x$ returned from $g(key, 11, 7)$ is $7$, then $f(key, 8)$ in Step 3 is applied to map $key$ to an integer $x \in \{0,1, \ldots, 7\}$ with equal probability given by
\[
P(X=x)=\dfrac{1}{8}, \quad x=0,1,\ldots,7.
\]
\\
Combining Steps 2 and 3 yields $P(X=x) = 1/11$ for $x = 0, 1, \ldots, 10$.  Thus, keys mapped to the range [11, 15] in Step 1 will be remapped uniformly over the range [0, 10].  This example shows how the algorithm maps keys to integers uniformly over the range [0, 10].  Probability distributions of hash values for functions $hash()$, $f()$, and $g()$ are generalized in Section IV. \\ \\
Also observe that $hash(key, 11)$ returns the result in Step 1 with probability $11/16$, and proceeds to Step 2 with probability $5/16$, because $f(key, 16)$ maps the key to an integer in the range [0, 15] with equal probability. The next section formalizes the algorithm, which can distribute keys among any number of buckets. \\
\section* {\pdfbookmark[1]{Power Consistent Hash Algorithm}{S3}\textbf{III.\quad Power Consistent Hash Algorithm}} 
\medskip
\noindent
In Equation 3.1, the function $hash(key,n)$ is constructed to map $key$ to an integer in the range $[0, n-1]$. The equation, consisting of  three cases, is computed using two auxiliary functions $f()$ and $g()$.  Let $n$ denote the number of buckets, and $m$ be the smallest power of 2 such that $m \ge n$.  We have $m/2 < n \leq m$. \\
\ \\
\textbf{Equation 3.1.}
\[
  hash(key, n) = 
    \begin{cases}
	r_{1} = f(key, m)  &  
\text{if }  r_{1} < n, \\
	r_{2}= g(key, n, m/2 - 1) & 
\text{if } r_{1} \geq n \text{ and } r_{2} > m/2 - 1, \\
	f(key, m/2) & 
\text{if } r_{1} \geq n \text{ and } r_{2} = m/2 - 1.
    \end{cases}
\]
\noindent 
\ \\ \\
The example in the previous section illustrated the case of $n=11$ and $m=16$, where $16$ is the smallest power of 2 greater than 11.  The hash spaces and distributions of hash values for $hash(), f()$ and $g()$ with given parameters are shown in Table 3.1. The weighted distribution for $g()$ is given by Property 4.3 in Section IV.  In this paper, functions may be denoted by trailing parentheses, such as $f()$ and $g()$, with parameters omitted.  \\ 
\definecolor{Gray}{gray}{0.9}
\setlength{\arrayrulewidth}{0.8pt}
\begin{table}[!h]
\begin{center}
\caption*{\textbf{Table 3.1 }  Hash Spaces and Distributions}
\begin{NiceTabular}{ wl{2.5cm}wl{2cm}wl{2cm} }
\hline
\rowcolor{Gray}
\rule{0pt}{12pt} \textit{Function} & \textit{Hash space} & \textit{Distribution} \\[3 pt]
\hline
\rule{0pt}{12pt} $hash(key,n)$ & $[0, n-1]$ & uniform \\ 
\rule{0pt}{9pt} $f(key,m)$ & $[0, m-1]$ & uniform \\ 
\rule{0pt}{9pt}  $f(key,m/2)$ & $[0,m/2-1]$ & uniform  \\ 
\rule{0pt}{9pt} $g(key,n,m/2-1)$ & $[m/2-1,n-1]$ & weighted \\[3 pt]
\hline
\end{NiceTabular}
\end{center}
\end{table}
\pagebreak
\begin{lstlisting}[caption={[abc]}]
$\textbf{\textsc{Power-Consistent-Hash}}$($key, n$)
1   $r1 = f(key,m)$  
2   if $r1 < n$
3      return $r1$
4   $r2 = g(key, n, m/2-1)$
5   if $r2 > m/2 -1$
6      return $r2$
7   return $f(key, m/2)$ 
\end{lstlisting}
\ \\
The \textsc{Power-Consistent-Hash} algorithm computes $hash(key, n)$ according to Equation 3.1 and Table 3.1.  It takes two parameters $key$ and $n$, evaluates the conditions, and returns the result of $r1, r2,$ or $f(key, m/2)$ accordingly.  $key$ is a numerical hash key.  $n$ denotes the number of buckets.  When $n$ is known, the corresponding $m$ is computed only once for all possible $key$.  Lines 1 and 7 call the same function $f()$ with different values on the second parameter. A unique characteristic of the algorithm is the following: (i) Given $n$ and $m$, the algorithm first maps the key with equal probability into a hash space of size $m$ (which is a power of 2); and (ii) optionally, it remaps the key into a smaller space of size $n$. Keys mapped to the range $[0, n-1]$ in (i) are skipped in (ii).  The operation is fast, consistent, and balanced.  In fact, the algorithm runs in $O(1)$ expected time, independent of the parameter $n$. \\ \\
The following theorems derived for Equation 3.1 are quite useful. The functions $hash()$,  $f()$ and $g()$ implemented by the algorithms possess those properties.\\ \\
\textbf{Theorem 3.1.} \enspace The function $hash(key,n)$ satisfies the distribution uniformity given by Property 4.1 if $f()$ and $g()$ used by $hash(key,n)$ satisfy the distribution properties given by Properties 4.2 and 4.3, respectively. \\ \\
\textbf{Theorem 3.2.} \enspace The function $hash(key,n)$ satisfies the mapping consistency defined by Property 5.1 if $f()$ and $g()$ used by $hash(key,n)$ satisfy the mapping consistency defined by Properties 5.2 and 5.3, respectively. \\ \\
These properties are given in the next two sections. Section VI will describe the algorithms to implement $f()$ and $g()$, each having $O(1)$ expected time. This yields an expected running time of $O(1)$ in \textsc{Power-Consistent-Hash}. Moreover, $f()$ and $g()$ implemented by the algorithms satisfy Properties 4.2 and 4.3, respectively. They also satisfy the mapping consistency defined by Properties 5.2 and 5.3, respectively. By the above theorems, $hash(key,n)$ implemented by \textsc{Power-Consistent-Hash} satisfies the distribution uniformity given by Property 4.1 as well as the mapping consistency defined by Property 5.1. \\
\section* {\pdfbookmark[1]{Distribution Properties}{S4}\textbf{IV.\quad Distribution Properties}} 
\medskip
\noindent
Properties 4.1 to 4.3 describe probability distributions of hash values for functions $hash()$, $f()$, and $g()$, respectively.  $hash()$ maps keys uniformly over a range.  It employs two auxiliary hash functions $f()$ and $g()$ to achieve distribution uniformity. $f()$ maps keys uniformly as described in Property 4.2, while $g()$ maps keys following a weighted probability distribution given by Property 4.3. Also $f()$ and $g()$ compute hash values independently. The distribution properties were illustrated in Section II.\\ \\
\noindent
\textbf{Property 4.1.} \enspace The function $hash()$ maps $key$ to an integer $x$ with equal probability given by $P(X=x)$.
\begin{align*}
hash(key, n) &= x, \\
P(X=x) &= \frac{1}{n}, \quad x = 0, 1, \ldots, n-1.
\end{align*}
\noindent
\ \\ \\
\textbf{Property 4.2.} \enspace The function $f()$ maps $key$ to an integer $x$ with equal probability  given by $P(X=x)$.  $m$ is a power of 2.
\begin{align*}
f(key, m) &= x, \\
P(X=x) &= \frac{1}{m},  \quad x = 0, 1, \ldots, m-1.
\end{align*}
\noindent
\\
\textbf{Property 4.3.} \enspace The function $g()$ maps $key$ to an integer $x$ with a weighted probability given by $P(X=x)$.  $s$ is an integer $\geq 0$.
\begin{align*}
g(key, n, s) &= x, \\ 
P(X=x) &=
\begin{cases}
\dfrac{s+1}{n}, &
x = s, \\ \\
\dfrac{1}{n}, &
x = s+1, \ldots, n-1.
\end{cases}
\end{align*}
\medskip 
\section* {\pdfbookmark[1]{Properties of Mapping Consistency}{S5}\textbf{V.\quad Properties of Mapping Consistency}}
\medskip
\noindent
For any given key, a hash function maps the key to an integer in a hash space, which is the set of all possible hash values of the function.  Under mapping consistency, key remapping is minimized when the hash space changes.\\ \\
\noindent
\textbf{Hash Space.} \enspace Given integer parameters $n_{1}, n_{2}, m_{1}, m_{2}$, $s$, individual hash spaces $S_{1}, S_{2}, \ldots, S_{6}$ are determined accordingly for all possible $key$ passed to the functions:
\begin{align*}
hash(key, n_{1})  & \in S_{1} = \{0,1, \ldots, n_{1}-1\}, \\
hash(key, n_{2})  & \in S_{2} = \{0,1, \ldots, n_{2}-1\}, \\
f(key, m_{1}) & \in S_{3} = \{0,1, \ldots, m_{1}-1\}, \\
f(key, m_{2}) &  \in S_{4} = \{0,1, \ldots, m_{2}-1\}, \\
g(key, n_{1}, s) &  \in S_{5} = \{s,s+1, \ldots, n_{1}-1\}, \\
g(key, n_{2}, s) &  \in S_{6} = \{s,s+1, \ldots, n_{2}-1\}.
\end{align*}
\ \\[-2 ex]
\noindent
\textbf{Example 5.1.}  \enspace This example illustrates Property 5.1.  In Table 5.1, there are 6 keys.  Columns $x$ and $y$ contain the hash values where the numbers of buckets are 10 and 19, respectively.  We have $x = y$ if $y < 10$.  Replacing 10 and 19 with $n_{2}$ and $n_{1}$, respectively, we then have $x=y$ if $y < n_{2} < n_{1}$. 
\definecolor{Gray}{gray}{0.9}
\setlength{\arrayrulewidth}{0.7pt}
\begin{table}[!h]
\begin{center}
\caption*{\textbf{Table 5.1 }  Example of Mapping Consistency}
\begin{NiceTabular}{ wc{1cm}wc{2.5cm}wc{2.5cm}}
\hline
\rowcolor{Gray}
\rule{0pt}{12pt} $key$ & $hash(key,10) = x$ & $hash(key,19) = y$ \\[3 pt]
\hline
\rule{0pt}{9pt} $k_{1}$ & 2 & 2 \\ 
\rule{0pt}{9pt} $k_{2}$ & 1 & 1  \\ 
\rule{0pt}{9pt} $k_{3}$  & 3 & 11  \\
\rule{0pt}{9pt} $k_{4}$  & 8 & 15 \\
\rule{0pt}{9pt} $k_{5}$  & 5 & 5  \\
\rule{0pt}{9pt} $k_{6}$  & 2 & 16 \\
\hline
\end{NiceTabular}
\begin{align*} 
hash(key, 10) = x, & \quad 0 \le x < 10, \\
hash(key, 19) = y, & \quad 0 \le y < 19. 
\end{align*}
\end{center}
\end{table}
\ \\
With this property, we can minimize key remapping and data redistribution when the number of buckets changes. The power consistent hash algorithm satisfies this property, and has the complexity of $O(1)$ space and $O(1)$ expected time. Similar examples can be derived for Properties 5.2 and 5.3.\\ \\
Properties 5.1 to 5.3 formalize mapping consistency for $hash()$, $f()$, and $g()$. Property 5.1 states that $hash(key, n_{1})$ and $hash(key, n_{2})$ map $key$ to the same integer $\alpha$ if $\alpha < n_{2} < n_{1}$. This property holds when the number of buckets changes from $n_{1}$ to $n_{2}$, and vice versa. 
In other words, key mapping does not change if $\alpha \in S_{2} \subset S_{1}$.  \\ \\
Similarly, in Property 5.2, $f(key, m_{1})$ and $f(key, m_{2})$ map $key$ to the same integer $\alpha$ if $\alpha < m_{2} < m_{1}$. Property 5.3 states that $g(key, n_{1}, s)$ and $g(key, n_{2}, s)$ map $key$ to the same integer $\alpha$ if $\alpha < n_{2} < n_{1}$. \\
\noindent
\ \\
\textbf{Property 5.1.} \enspace Suppose that for any $key$ and integer $n_{1}$,
\begin{alignat*}{2}
hash(key, n_{1})  &= \alpha,  \quad &&\alpha \in \{0,1, \ldots, n_{1}-1\}. \\[-1ex]
\intertext{Then} \\[-5ex]
hash(key, n_{2}) &= \alpha,  \quad &&\text{for all }n_{2} \text{ such that }  \alpha < n_{2} < n_{1}.
\end{alignat*}
\ \\
\noindent
\textbf{Property 5.2.} \enspace Suppose that for any $key$, and $m_{1}$ that is a power of 2,
\begin{alignat*}{2}
f(key, m_{1})  &= \alpha,  \quad &&\alpha \in \{0,1, \ldots, m_{1}-1\}.  \\[-1ex]
\intertext{Then}\\[-5ex]
f(key, m_{2})  &= \alpha, \quad &&\text{for all }m_{2} \text{ such that } \alpha < m_{2} < m_{1}, \text{ and}\\
&\phantom{= x} \quad && m_{2} \text{ is a power of 2}.
\end{alignat*}
\noindent 
\textbf{Property 5.3.} \enspace Suppose that for any $key$, and integers $n_{1}, s,$ where $s \geq 0$,
\begin{alignat*}{2}
g(key, n_{1},s)  &= \alpha,  \quad &&\alpha \in \{s,s+1, \ldots, n_{1}-1\}.  \\[-1ex]
\intertext{Then}\\[-5ex]
g(key, n_{2},s) &= \alpha, \quad &&\text{for all }n_{2} \text{ such that } \alpha < n_{2} < n_{1}.
\end{alignat*}

\medskip
\section*{\pdfbookmark[1]{Algorithms for Auxiliary Functions}{S6}\textbf{VI.\quad Algorithms for Auxiliary Functions}}
\medskip
\noindent
Algorithm-f and Algorithm-g compute $f(key, m)$ and $g(key, n, s)$, respectively.  $f(key, m)$ maps keys uniformly over a range, while $g(key, n, s)$ maps keys following a weighted probability distribution given by Property 4.3.
\subsection*{\pdfbookmark[2]{Algorithm-f}{S6A}\textit{A.\quad Algorithm-f}}
\noindent
Algorithm-f computes $f(key, m)$ for the input parameters $key$ and $m$, where $key$ is a numerical hash key with at least $\log_{2}(m)$ bits, and $m$ must be a power of 2.  Assume $key$ contains bits that are reasonably random.  The algorithm returns an integer in the range $[0, m-1]$ with equal probability.  The \textsc{Power-Consistent-Hash} algorithm in Section III calls $f(key, m)$ once, and $f(key, m/2)$ at most once.\\
\begin{lstlisting}
$\textbf{Algorithm-f}$($key, m$)
1   $kBits = (key\enspace \& \enspace (m-1))$  
2   if $kBits == 0$
3      return $0$
4   $j$ = $\textsc{FindLastOneBit}$($kBits$)
5   $h = 1 << j$
6   $r = h + $($\textsc{Rand}$($key, j$)$\enspace \& \enspace (h-1))$
7   return $r$ 
\end{lstlisting}
\noindent
\ \\
Line 1 in the pseudocode extracts $\log_{2}(m)$ bits from a given $key$.  `\&' in lines 1 and 6 denotes the bitwise-AND operator.  In line 4, \textsc{FindLastOneBit} returns the bit index of the most significant bit set to 1 in $kBits$, which can be done in $O(1)$ time using a hardware instruction (such as BSR, LZCNT \cite{AM}) or lookup table. The bit index is an unsigned offset from bit 0.  Line 5 computes $2^j$ by using the bitwise left shift operator `$<<$'. Line 6 produces a random integer $r$ in the range $[h, 2h-1]$ with equal probability.  \textsc{Rand}($key,j$) returns a pseudo-random integer deterministically based on the values of $key$ and $j$. Additionally, values returned from \textsc{Rand}($key, j$) are reasonably random and independent for distinct pairs of $key$ and $j$.  \\ \\
Overall, Algorithm-f runs in $O(1)$ time, independent of the input parameter $m$.  Moreover, the algorithm satisfies the distribution uniformity given by Property 4.2 and ensures the mapping consistency defined by Property 5.2. Also note that Algorithm-f can be used as a standalone consistent hash method when the number of buckets is always a power of 2. \\
\definecolor{Gray}{gray}{0.9}
\setlength{\arrayrulewidth}{0.7pt}
\begin{table}[!h]
\begin{center}
\caption*{\textbf{Table 6.1 }  Sample Values from Algorithm-f}
\begin{NiceTabular}{ wc{0.3cm}wc{1.4cm}wl{0.5cm}wl{0.3cm}wc{1.5cm}|wl{4cm} }
\hline
\rowcolor{Gray}
\rule{0pt}{12pt} $key$ & $kBits$ & $j$ & $h$ & $f(key,16)$ & \textit{Range: {$[h, 2h-1]$}} \\[3 pt]
\hline
\rule{0pt}{9pt} $k_{1}$ & 0001 & 0 & 1 & $R(k_{1}, 0)$ & 1 \\ 
\rule{0pt}{9pt} $k_{2}$ & 0010 & 1 & 2 & $R(k_{2}, 1)$ & 2, 3 \\ 
\rule{0pt}{9pt} $k_{3}$ & 0101 & 2 & 4 & $R(k_{3}, 2)$ & 4, 5, 6, 7 \\
\rule{0pt}{9pt} $k_{4}$ & 1100 & 3 & 8 & $R(k_{4}, 3)$ & 8, 9, 10, 11, 12, 13, 14, 15\\
\hline
\end{NiceTabular}
\end{center}
\end{table}
\ \\
Table 6.1 shows sample values of $kBits, j$, and $h$ in the algorithm for input parameters $key$ and $m=16$.  There are four different keys.  For $m=16$, the algorithm extracts 4 ($=\log_{2}16$) bits from $key$ and stores in $kBits$.  The second column contains the least significant 4 bits of $kBits$.  Column $j$ shows the bit index of most significant bit set to 1 in $kBits$.  Column $h$ has the value of $2^{j}$. Based on the values of $key$ and $j$, the algorithm produces a random integer $r$ in the corresponding range $[h, 2h-1]$ as the result.  The integer $r$ produced is denoted by $R(key, j)$ in column $f(key, 16)$.  A special case occurs when $kBits$ equals 0, for which the algorithm returns 0.  Assuming the 4 bits of $kBits$ are random bits, the algorithm returns an integer in the range [0, 15] with equal probability.  \\ \\
As an example to illustrate Property 5.2, suppose $f(k_{2},16)=3$.  We have 
\[
f(k_{2},16) = R(k_{2}, j) = 3, \text{ where } j = 1.
\]
Then 
\[
f(k_{2},8)=f(k_{2},4)=R(k_{2}, j) = 3,
\]
since the value of $j$ is still 1 when $m=8$ or 4.  Therefore $f(k_{2}, 16)=f(k_{2}, m)$ for $m= 8, 4$, 
where $f(k_{2}, 16) < m < 16$, as indicated in Property 5.2. 
\subsection*{\pdfbookmark[2]{Algorithm-g}{S6B}\textit{B.\quad Algorithm-g}}
\noindent
Algorithm-g computes $g(key, n, s)$, and returns an integer in the range $[s, n-1]$ with a weighted probability given by Property 4.3.  It generates a sequence of pseudo-random integers in increasing order and deterministically based on the values of $key, n, s$.  The sequence starts with integer $s, s \geq 0$, and is upper bounded by $n-1$. Let $A_{r}$ denote the event that an integer $r$ occurs in the sequence.  The probability of $A_{r}$ is given by
\[
P(A_{r}) =
\begin{cases}
1, &
r = s, \\ \\
\dfrac{1}{r+1}, &
r = s+1, \ldots, n-1. \\
\end{cases}
\]
\ \\
The algorithm returns the last integer in the sequence as the result of $g(key, n, s)$. Assume the events $A_{r}$ for all $r$ are independent.  By calculating the probability of $r$ being the last integer in the sequence, we obtain the weighted distribution in Property 4.3.  That is, keys are distributed according to Property 4.3 as follows:
\begin{align*}
g(key, n, s) &= x, \\ 
P(X=x) &=
\begin{cases}
\dfrac{s+1}{n}, &
x = s, \\ \\
\dfrac{1}{n}, &
x = s+1, \ldots, n-1.
\end{cases}
\end{align*}
\ \\
To generate $r$ sequentially based on $P(A_{r})$, Algorithm-g runs a loop to generate random integers starting with $s, s \geq 0$.  Let $x$ denote the random integer generated in the current iteration.  Initially, $x$ is set to the value of $s$.  The next random integer is computed as follows:
\begin{flalign*}
&\text{\qquad 1.\quad Generate }U.  &\\
&\text{\qquad 2.\quad Compute }r = \min \left\{ j \, : \, U > \dfrac{x+1}{j+1} \right\}. &\\
&\text{\qquad 3.\quad Set }x = r \text{ if } r < n. &\\[-6ex]
\end{flalign*}
\ \\
$U$ denotes the next random number from a generator $U(0,1)$ that generates random numbers uniformly over range (0, 1) and deterministically based on the given key.  Step 2 determines the smallest integer $j$ satisfying the inequality.  If $r < n$, the algorithm sets $x$ to the value of $r$ and repeats steps 1 to 3.  Otherwise, the algorithm returns the current value of $x$ as the result.  Each iteration of the loop takes $O(1)$ time to generate a random integer denoted by $x$. \\ \\
Next, we show that the computation of $g(key, n, s)$ has $O(1)$ expected time if $n < 2(s+1)$.  Based on the probability $P(A_{r})$ for all $r$, the expected number of iterations is given by
\begin{align*}
&1+\frac{1}{s+2}+\frac{1}{s+3}+\cdots+\frac{1}{n} \\
<\enspace&1+\int_{s+1}^{n} \frac{dx}{x}\\
=\enspace&1+\ln n - \ln (s+1).
\end{align*}
Letting $n < 2(s+1)$ yields
\begin{align*}
&1+\ln n - \ln (s+1) \\
<\enspace&1+\ln (2(s+1)) - \ln(s+1) \\
=\enspace&1+\ln 2 \\
<\enspace&1.7.
\end{align*}
The expected number of iterations is thus reduced to a constant less than 1.7.  The \textsc{Power-Consistent-Hash} algorithm in Section III calls $g(key, n, s)$ with $ s = (m/2-1)$, where $m$ is the smallest power of 2 greater than $n$.  Since $n < m = 2(s+1)$, the expected time to compute $g(key, n, m/2-1)$ is $O(1)$ independent of $n$.  With high probability, Algorithm-g only needs to generate a very short sequence of random integers when starting with $s=(m/2-1)$, and returns the last one in the sequence.  The length of the sequence also has a very small variance $\sigma^2 < \ln 2$. Thus, the upper bound for computing $g(key, n, m/2-1)$ is essentially a small constant. In practice, Algorithm-g can set a proper limit on the number of iterations, which should have a negligible effect on the distribution of hash values. \\ \\
Algorithm-g satisfies Property 5.3, which can be illustrated with an example.  Suppose $g(key, 15, 7)$ $= 9$.  By Property 5.3, we have $g(key, 12, 7) = 9$ since $9 < 12 < 15$. To see this, observe that the algorithm generates random integers sequentially and deterministically.  It returns the largest random integer less than $n$ as the result.  Let $x, r$ denote two adjacent random integers. If $g(key, 15, 7) = 9$, the algorithm must stop at $x=9$, where $r \ge 15$.  Similarly, when computing $g(key, 12, 7)$, it must stop at $x=9$ because $9 < 12$ and $r \ge 15$.  Thus $g(key, 12, 7) = 9$.  This demonstrates Property 5.3. 
\subsection*{\pdfbookmark[2]{Summary of Time Complexity}{S6C}\textit{C.\quad Summary of Time Complexity}}
\noindent
The \textsc{Power-Consistent-Hash} algorithm in Section III calls $f(key, m)$ once, and calls $f(key$, $m/2)$ and $g(key, n, m/2-1)$ each at most once for input parameters $key$ and $n$, where $m$ is the smallest power of 2 such that $ m \ge n$. This gives an expected running time of $O(1)$ independent of $n$.  In comparison, jump consistent hash \cite{LV} executes a loop that iterates from bucket number 0, which has $O(\ln n)$ expected time.  The next section provides performance test results.\\
\begin{figure}[h!]
\centering
\includegraphics[scale=1.0]{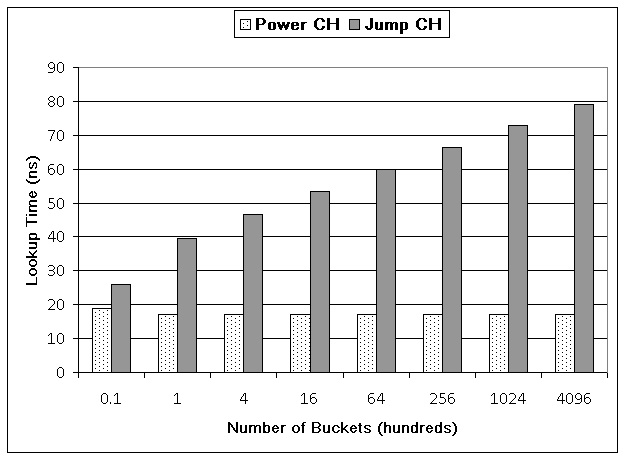}
\caption*{\textbf{Figure 2 } \enspace Evaluation of  Lookup Time}
\end{figure}
\section*{\pdfbookmark[1]{Performance Evaluation}{S8}\textbf{VII.\quad Performance Evaluation}}
\medskip
\noindent
Performance testing has been conducted to compare power consistent hash with jump consistent hash \cite{LV}.  In 
Figure 2,  X-axis is the number of hash buckets.  Y-axis is the average lookup time.  Lookup time stays flat in power consistent hash (Power CH).  In contrast, lookup time grows with the number of buckets in jump consistent hash (Jump CH).  The growth rates agree with the asymptotic analysis of the algorithms: $O(1)$ in power consistent hash versus $O(\ln n)$ in jump consistent hash.  Power consistent hash is much faster with superior scalability.
\medskip
\section*{\pdfbookmark[1]{Rehashing in Constant Time}{S7}\textbf{VIII.\quad Rehashing in Constant Time}}
\medskip
\noindent
Fast lookup in $O(1)$ expected time also has the unique advantage of fast rehashing over other consistent hash algorithms that have higher running time.  When a bucket is not available, rehashing can be used to map affected keys to other buckets.  For a given key, the hash algorithm returns an integer $x$ in the range [0, $n-1$], which can be resized from the upper end.  Suppose bucket $x$ is unavailable or removed and $x$ is not at the upper end of the range. Then rehashing can map the key to another integer in the same range. Rehashing can be repeated to map the key to an available bucket.\\ \\
Rehashing also has $O(1)$ expected time. The algorithm probes randomly and iteratively to find an available bucket.  It stops rehashing once it finds an available bucket or picks a fallback bucket.  The likelihood of fallback is exponentially reduced with respect to the number of times rehashing is performed.  Therefore, it is sufficient to reserve a small portion of storage capacity for fallback buckets while keeping $O(1)$ lookup time.  This ensures that the number of times rehashing is performed is at most a small constant in the worst case.\\ \\
The rehashing technique described is simpler, faster, and more scalable than alternatives that need to maintain a list of available buckets.  In particular, the rehashing technique in $O(1)$ expected time can solve two types of problems efficiently: (i) if a bucket is unavailable, it can map affected keys to some other buckets in real time, minimizing system disruption; and (ii) if a bucket is overloaded, the technique can reduce the load on demand by remapping a fraction of the keys currently mapped to the bucket to some other buckets.  In a distributed environment, the algorithm can run on multiple servers for high throughput and availability. Those servers share minimal global states that need to be maintained consistently, thereby achieving high autonomy and fast actions upon change.
\medskip
\section*{\pdfbookmark[1]{Conclusion}{S9}\textbf{IX.\quad Conclusion}}
\medskip
\noindent
The power consistent hash (Power CH)  algorithm computes a hash function using two auxiliary hash functions to achieve $O(1)$ expected time for key lookup.  Keys are uniformly distributed among buckets, as reflected in Equation 3.1. With the complexity of $O(1)$ space and $O(1)$ expected time, it can support many use cases where the number of buckets can dynamically change. Performance testing shows superior scalability.  Moreover, it satisfies the mapping consistency property for which key remapping is minimized when the number of buckets changes.  Lastly, when a bucket is unavailable or overloaded, a rehashing technique can map each of affected keys to another bucket in $O(1)$ expected time. \\
\pdfbookmark[1]{Note}{S10}
\section*{\textbf{Note}}
\noindent
The invention and major ideas in this paper are documented in U.S. Patent No. 11,429,452, granted on August 30, 2022. \cite{LE}\\
\pdfbookmark[1]{References}{S11}
\renewcommand\refname{\textbf{References}}
\hypersetup{bookmarksdepth=-2}
\bibliographystyle{IEEEtran}

\end{document}